\documentclass[10pt,conference]{IEEEtran}
\IEEEoverridecommandlockouts
\newcommand{\para}[1]{\vspace{2pt}\noindent\textbf{#1.~}}
\newcommand{\ignore}[1]{}
\newcommand{\system}{\textit{\sloppy{SmartReco}\@}}

\usepackage[dvipsnames]{xcolor}
\usepackage{cite}
\usepackage{amsmath,amssymb,amsfonts}
\usepackage{algorithmic}
\usepackage{graphicx}
\usepackage{textcomp}
\usepackage{xcolor}
\def\BibTeX{{\rm B\kern-.05em{\sc i\kern-.025em b}\kern-.08em
    T\kern-.1667em\lower.7ex\hbox{E}\kern-.125emX}}

\usepackage{algorithm}
\usepackage{tcolorbox}
\usepackage{listings, xcolor}

\definecolor{verylightgray}{rgb}{.97,.97,.97}

\lstdefinelanguage{Solidity}{
	keywords=[1]{anonymous, assembly, assert, break, call, callcode, case, catch, class, constant, continue, constructor, contract, debugger, default, delegatecall, delete, do, else, emit, event, experimental, export, external, false, finally, for, function, gas, if, implements, import, in, indexed, instanceof, interface, internal, is, length, library, log0, log1, log2, log3, log4, memory, modifier, new, payable, pragma, private, protected, public, pure, push, require, return, returns, revert, selfdestruct, send, solidity, storage, struct, suicide, super, switch, then, this, throw, transfer, true, try, typeof, using, value, view, while, with, addmod, ecrecover, keccak256, mulmod, ripemd160, sha256, sha3}, % generic keywords including crypto operations
	keywordstyle=[1]\color{blue}\bfseries,
	keywords=[2]{address, bool, byte, bytes, bytes1, bytes2, bytes3, bytes4, bytes5, bytes6, bytes7, bytes8, bytes9, bytes10, bytes11, bytes12, bytes13, bytes14, bytes15, bytes16, bytes17, bytes18, bytes19, bytes20, bytes21, bytes22, bytes23, bytes24, bytes25, bytes26, bytes27, bytes28, bytes29, bytes30, bytes31, bytes32, enum, int, int8, int16, int24, int32, int40, int48, int56, int64, int72, int80, int88, int96, int104, int112, int120, int128, int136, int144, int152, int160, int168, int176, int184, int192, int200, int208, int216, int224, int232, int240, int248, int256, mapping, string, uint, uint8, uint16, uint24, uint32, uint40, uint48, uint56, uint64, uint72, uint80, uint88, uint96, uint104, uint112, uint120, uint128, uint136, uint144, uint152, uint160, uint168, uint176, uint184, uint192, uint200, uint208, uint216, uint224, uint232, uint240, uint248, uint256, var, void, ether, finney, szabo, wei, days, hours, minutes, seconds, weeks, years},	% types; money and time units
	keywordstyle=[2]\color{teal}\bfseries,
	keywords=[3]{block, blockhash, coinbase, difficulty, gaslimit, number, timestamp, msg, data, gas, sender, sig, value, now, tx, gasprice, origin},	% environment variables
	keywordstyle=[3]\color{violet}\bfseries,
	identifierstyle=\color{black},
	sensitive=true,
	comment=[l]{//},
	morecomment=[s]{/*}{*/},
	commentstyle=\color{gray}\ttfamily,
	stringstyle=\color{red}\ttfamily,
	morestring=[b]',
	morestring=[b]"
}

\usepackage{graphicx}
\usepackage{subcaption}
\usepackage{url}
\usepackage{hyperref}
\usepackage{booktabs}
\usepackage{multirow}
\graphicspath{{picture}}
\usepackage{enumitem}
\renewcommand{\algorithmicrequire}{ \textbf{Input:}}
\renewcommand{\algorithmicensure}{ \textbf{Output:}}
\begin{document}

\title{SmartReco: Detecting Read-Only Reentrancy via Fine-Grained Cross-DApp Analysis}

\author{\IEEEauthorblockN{Jingwen Zhang\IEEEauthorrefmark{2}\IEEEauthorrefmark{4},
Zibin Zheng\IEEEauthorrefmark{2}\IEEEauthorrefmark{6},
Yuhong Nan\thanks{* Yuhong Nan is the corresponding author.}\IEEEauthorrefmark{2}\IEEEauthorrefmark{6}\IEEEauthorrefmark{1}, 
Mingxi Ye\IEEEauthorrefmark{2}\IEEEauthorrefmark{6},
Kaiwen Ning\IEEEauthorrefmark{2}\IEEEauthorrefmark{4},
Yu Zhang\IEEEauthorrefmark{3}\IEEEauthorrefmark{4}, and
Weizhe Zhang\IEEEauthorrefmark{3}\IEEEauthorrefmark{4}}
\IEEEauthorblockA{\IEEEauthorrefmark{2}Sun Yat-sen University, \href{mailto:zhangjw273@mail2.sysu.edu.cn,yemx6@mail2.sysu.edu.cn,ningkw@mail2.sysu.edu.cn}{\{zhangjw273, yemx6, ningkw\}@mail2.sysu.edu.cn}, \href{mailto:zhzibin@mail.sysu.edu.cn,nanyh@mail.sysu.edu.cn}{\{zhzibin, nanyh\}@mail.sysu.edu.cn}\\
\IEEEauthorrefmark{3}Harbin Institute of Technology, \href{mailto:yuzhang@hit.edu.cn,wzzhang@hit.edu.cn}{\{yuzhang, wzzhang\}@hit.edu.cn}\\
\IEEEauthorrefmark{4}Peng Cheng Laboratory, \IEEEauthorrefmark{6}GuangDong Engineering Technology Research Center
of Blockchain
}}

% \author{\IEEEauthorblockN{1\textsuperscript{st} Jingwen Zhang}
% \IEEEauthorblockA{\textit{School of Software Engineering} \\
% \textit{Sun Yat-sen University}\\
% Zhuhai, China \\
% zhangjw273@mail2.sysu.edu.cn}
% \and
% \IEEEauthorblockN{2\textsuperscript{nd} Zibin Zheng}
% \IEEEauthorblockA{\textit{School of Software Engineering} \\
% \textit{Sun Yat-sen University}\\
% Zhuhai, China \\
% zhzibin@mail.sysu.edu.cn}
% \and
% \IEEEauthorblockN{3\textsuperscript{rd} Yuhong Nan\IEEEauthorrefmark{*}\thanks{* Yuhong Nan is the corresponding author}}
% \IEEEauthorblockA{\textit{School of Software Engineering} \\
% \textit{Sun Yat-sen University}\\
% Zhuhai, China \\
% nanyh@mail.sysu.edu.cn}
% \and
% \IEEEauthorblockN{4\textsuperscript{th} Mingxi Ye}
% \IEEEauthorblockA{\textit{School of Software Engineering} \\
% \textit{Sun Yat-sen University}\\
% Zhuhai, China \\
% yemx6@mail2.sysu.edu.cn}
% \and
% \IEEEauthorblockN{5\textsuperscript{th} Kaiwen Ning}
% \IEEEauthorblockA{\textit{School of Software Engineering} \\
% \textit{Sun Yat-sen University}\\
% Zhuhai, China \\
% ningkw@mail2.sysu.edu.cn}
% \and
% \IEEEauthorblockN{6\textsuperscript{th} Yu Zhang}
% \IEEEauthorblockA{\textit{Harbin Institute of Technology}\\
% Harbin, China \\
% yuzhang@hit.edu.cn}
% \and
% \IEEEauthorblockN{7\textsuperscript{th} Weizhe Zhang}
% \IEEEauthorblockA{\textit{Harbin Institute of Technology}\\
% Harbin, China \\
% wzzhang@hit.edu.cn}
% }

\maketitle

\begin{abstract}
Despite the increasing popularity of Decentralized Applications (DApps), they are suffering from various vulnerabilities that can be exploited by adversaries for profits. Among such vulnerabilities, Read-Only Reentrancy (called ROR in this paper), is an emerging type of vulnerability that arises from the complex interactions between DApps. In the recent three years, attack incidents of ROR have already caused around 30M USD losses to the DApp ecosystem. Existing techniques for vulnerability detection in smart contracts can hardly detect Read-Only Reentrancy attacks, due to the lack of tracking and analyzing the complex interactions between multiple DApps.

In this paper, we propose \system{}, a new framework for detecting Read-Only Reentrancy vulnerability in DApps through a novel combination of static and dynamic analysis (i.e., fuzzing) over smart contracts. The key design behind \system{} is threefold: (1) \system{} identifies the boundary between different DApps from the heavy-coupled cross-contract interactions.  (2) \system{} performs fine-grained static analysis to locate points of interest (i.e., entry functions) that may lead to ROR. (3) \system{} utilizes the on-chain transaction data and performs multi-function fuzzing (i.e., the entry function and victim function) across different DApps to verify the existence of ROR.
%
%Unlike traditional cross-contract analysis, \system{} uses DApp information to achieve more fine-grained data collection. With the collected data, \system{} employs a combination of dynamic and static methods to detect Read-only reentrancy. In particular, instead of manually gathering contract addresses, \system{} utilizes a novel DApp collection algorithm, effectively solving the challenges of information gathering from DApps. 
%
Our evaluation of a manual-labeled dataset with 45 RORs shows that \system{} achieves a precision of 88.64\% and a recall of 86.67\%. In addition, \system{} successfully detects 43 new RORs from 123 popular DApps. The total assets affected by such RORs reach around 520,000 USD.

%The evaluation of \system{} on 123 DApps demonstrates its effectiveness in identifying 31 Read-only reentrancy vulnerabilities with a precision of 91\%. \jw{Among these 31 contracts, \system{} successfully detects 5 unreported vulnerabilities.} When compared to existing tools, while they fail to identify any Read-only reentrancy, \system{} exhibits lower false positives.
\end{abstract}


\section{Introduction}
Decentralized Applications (DApps) are applications including multiple smart contracts, which are code snippets that contain multiple functions to accomplish specific functionalities and can be executed on the blockchain.
% Due to their anonymity and immutability, DApps are widely used in digital collectibles~\cite{nftsok}, financial service~\cite{defisok}, and gaming~\cite{gamesok}. 
Due to the inherent financial property of DApps, the security of DApps is extremely important. For example, in recent years, the entire DApp ecosystem has suffered from various types of attacks, such as front-running~\cite{kolluri2019exploiting}, price manipulation~\cite{Defiranger}, and access control~\cite{ethainter}
, resulting in billions of dollars in losses~\cite{defiattacksok}.

\para{Reentrancy and Read-Only Reentrancy}
As one of the most typical vulnerabilities in smart contracts, the reentrancy vulnerability is leveraged to manipulate global states, such as the contract's state, to make the contract's behavior inconsistent with expectations~\cite{reguard}. In recent years, with the increasing attention on smart contract security, reentrancy has been addressed by many previous research~\cite{icychecker, confuzzius, oyente,deeplearningReentrancy}.

Read-Only Reentrancy (ROR), a new type of reentrancy vulnerability first reported in 2022~\cite{Marketxyz}, is a cross-DApp attack that specifically exploits functions in different DApps’ contracts. The key difference between ROR and traditional reentrancy is that ROR takes place between the smart contracts of independent DApps, while traditional reentrancy performs within the smart contract(s) of a single DApp.
To exploit ROR, the adversary first manipulates a specific state of one DApp by invoking one of its function (i.e., the \textit{entry function}). Then, the adversary performs the attack by hijacking the execution control flow and invoking another DApp's function (i.e., the \textit{victim function}) which relies on the aforementioned state, like calculating the token price with the token supply (see Section~\ref{subsec:read-only} for more details). 
Right now, ROR has already resulted in losses exceeding 30M USD and poses a threat of more than 100M USD. Due to the invisibility and potential risks of ROR, it has been selected as one of the top ten blockchain hacking techniques in 2022~\cite{2022top10}.

\para{Challenges in detecting Read-Only Reentrancy}
Despite a number of recent research focusing on detecting smart contract vulnerabilities, detecting ROR based on existing techniques is by no means trivial. Specifically, there are three fundamental challenges as we elaborate below.

Firstly, identifying the boundaries between DApps is difficult. Specifically, as ROR is a cross-DApp attack, detecting ROR requires analyzing complex relationships between DApps. However, the anonymity of the blockchain prevents access to smart contract identification, such as which DApp a smart contract belongs to. Prior research~\cite{disentanglingDefi,defiIntegration, icychecker} simply look up the contract address from the DApp's official website, or query third-party API (e.g., DAppRadar~\cite{dappRadar} and Thegraph~\cite{theGraph}). However, this information is often incomplete, as new smart contracts included by the DApp are not updated in a timely fashion.

%Prior research~\cite{disentanglingDefi, icychecker} merely relies on the list of contract addresses released by DApp developers to determine the DApp boundary. This, however, can easily overlook new smart contracts that are not released by DApp developers in a timely fashion. 

%\yuhong{With such an inaccurate DApp-contract mapping information, contract-level vulnerability detection tools, such as xFuzz~\cite{xFuzz} and SmartDagger~\cite{smartDagger}, will introduce a large number of false positives due to the inclusion of irrelevant contracts}.

%With such an inaccurate DApp-contract mapping information, contract-level vulnerability tools, such as xFuzz~\cite{xFuzz} and SmartDagger~\cite{smartDagger}, can only analyze at contract level and introduce a significant number of irrelevant contracts when detecting ROR, resulting in high false positives.

% Firstly, it is hard to understand the boundaries between DApps. Specifically, traditional reentrancy only occurs within a single DApp, while ROR is a cross-DApp attack during interactions between two or more DApps. Therefore, detecting ROR requires analyzing complex relationships between DApps. However, traditional cross-contract analysis, such as xFuzz~\cite{xFuzz} and SmartDagger~\cite{smartDagger}, can only analyze at contract level and introduces a significant number of irrelevant contracts when detecting ROR, resulting in high false positives.

Secondly, identifying the entry function (the start point) and the victim function (the end point) of ROR is quite challenging, given the unlimited number of DApps and their publicly accessible functions. Specifically, a function within a DApp can be called by anyone and any other DApp, resulting in a large search space. Moreover, these two functions do not share any explicit dependency (e.g., a shared state), or within the same call chain. 
Existing methods such as IcyChecker~\cite{icychecker} and DeFiTainter~\cite{defiTainter} blindly consider all functions as the candidates, which is not feasible for detecting ROR due to the huge exploration space. 

%Lastly, it is still rather difficult to verify RORs given the entry function and victim function, as the ROR path is sophisticatedly designed. 
Lastly, even with the given entry function and victim function pairs, it is still difficult to find a valid path (call chain) that can trigger ROR. On the one hand, static analysis methods, such as Pluto~\cite{pluto} and Sailfish~\cite{sailfish}, can not recover the call chain of ROR, due to the lack of concrete runtime contextual information. 
On the other hand, fuzzing-based methods, such as ConFuzzius~\cite{confuzzius} and ContractFuzzer~\cite{contractFuzzer}, randomly generate fuzzing seeds for multi-function (e.g., entry function and victim function) in different DApps. However, these approaches can hardly simulate the actual cross-DApp context (e.g., simultaneously bypassing the internal checks in both DApps ), not to mention triggering the potential RORs.

\para{Our work}
In this paper, we propose \system{}, a new framework to detect Read-Only Reentrancy vulnerabilities in smart contracts. To address the aforementioned challenges, \system{} comes with the following three unique designs: (1) cross-DApp analysis to identify DApp boundaries, (2) fine-grained static analysis for locating potential entry functions, (3) multi-function fuzzing across different DApps for effectively verifying RORs.

More specifically, the cross-DApp analysis aims to incorporate accurate DApp-based contextual data to identify valid attack surfaces. Here, the DApp-based contextual data refers to records of the interactions between DApps during transaction execution, such as cross-DApp call and cross-DApp state read and write. To facilitate this analysis, we propose an enhanced DApp boundary identification method based on DApp (contract) builders (Section~\ref{subsec:DApp_identification}). Our observation is that a DApp is usually managed by a fixed group of accounts as the builders, allowing us to more accurately identify DApp boundaries. With this method, we build the first DApp builder dataset $D_{builder}$ (with 334 unique builders). To the best of our knowledge, this is the first approach and dataset for accurately classifying smart contracts for DApps.

Based on the collected DApp information, \system{} performs fine-grained static analysis to find potential entry functions. Although there is no explicit dependency between the entry points and the victim points of ROR, the entry points must be associated with specific cross-DApp invocations along the call chain of a ROR. Therefore, with DApp-based contextual data, \system{} analyzes all cross-DApp function invocations on the call chain and identifies potential ROR entry points. (Section~\ref{subsec:data_collection} and Section~\ref{subsec:cross_dapp_analysis}).

% with the help of ROR contextual data such as state dependencies, 

Then, to verify whether the potential ROR entry points can trigger ROR, \system{} adopts a multi-function fuzzing mechanism to capture more accurate context information. To detail, \system{} generates inputs for functions in different DApps based on historical transactions, as such transactions provide a more realistic scenario for finding the valid paths and contexts to invoke ROR (Section~\ref{subsec:verification}).

To evaluate the effectiveness of \system{}, we construct a ROR dataset with 45 attack instances based on publicly available attack reports. We manually check and label the contracts (functions) related to these attack incidents. To the best of our knowledge, the dataset covers all reported RORs related to smart contracts as of Mar. 2024 (see Section~\ref{subsec:Implementation} for more details). 
The evaluation results show that \system{} has a precision of 88.64\% and a recall of 86.67\% over this dataset. In the meantime, two state-of-the-art frameworks, ityFuzz~\cite{ityFuzz} and Sailfish~\cite{sailfish}, can not detect any RORs in the dataset. 
Besides, with the help of \system{}, we perform an in-the-wild ROR inspection over 123 most popular DApps (with 2,676 smart contracts). Indeed, \system{} successfully detects 43 unreported RORs. The vulnerabilities discovered by \system{} impact approximately 520,000 USD.

%In addition, with an evaluation of 123 popular DApps, which contain 2,676 smart contracts, \system{} successfully detects 43 unreported RORs. Overall, the vulnerabilities discovered by \system{} impact approximately 520,000 USD worth of assets.
%
% In the meantime, we also compare the effectiveness of \system{} with ityFuzz~\cite{ityFuzz} and Sailfish~\cite{sailfish}, two state-of-the-art tools in detecting reentrancy. Results show \system{} significantly outperforms these tools (see Section~\ref{chap:evaluation}). 

To promote smart contract security development, we have open-sourced \system{} and the datasets used in our work~\footnote{https://github.com/zzz-sysu/smartReco}. In summary, this paper makes the following contributions:
\begin{itemize}
    \item We propose \system{}, a novel framework for detecting ROR - an emerging type of reentrancy vulnerabilities in smart contracts. To the best of our knowledge, \system{} is the first research to detect Read-Only Reentrancy.
    \item We design a series of new mechanisms, such as identifying DApp boundaries, and collecting DApp-based contextual data, to facilitate fine-grained cross-DApp analysis.
    \item We perform extensive experiments to verify the effectiveness of \system{}. The results indicate that \system{} can detect new RORs while maintaining lower false positives and false negatives.
    \item We release the artifact and the datasets of \system{}.
\end{itemize}

The rest of the paper is organized as follows: Section~\ref{sec:background} provides the background of ROR and the motivation of our research. Section~\ref{chap:design} highlights the challenges and solutions for ROR detection. Section~\ref{chapt:approach_details} elaborates on the implementations of \system{}. Section~\ref{chap:evaluation} introduces the experimental setup and evaluation. Section~\ref{sec:discussion} gives more discussions about ROR and the limitations of \system{}. Section~\ref{chap:related_work} presents related work and Section~\ref{chap:conclusion} concludes the paper.

\section{Background and Motivation}
\label{sec:background}

\subsection{Smart Contract and Contract Execution}
\label{subsec:background}
Decentralized Application (DApp) typically consists of a UI frontend and a backend that uses smart contracts to store and process data~\cite{metcalfe2020ethereum}. Smart contracts are codes that can be executed on blockchain platforms (such as Ethereum~\cite{ethereum} and BSC Chain~\cite{bnb}) and contain multiple functions to accomplish specific functionalities, e.g., transferring funds. To execute smart contracts, it is necessary to first compile the smart contract into bytecode and deploy it on the blockchain. In blockchain, there are two roles: user (called externally owned account, EOA) and smart contract, and they are both distinguished by a unique identifier called address.

%\para{Smart contract execution}
Users and smart contracts can send external and internal transactions respectively to a contract address to invoke functions and the contract is responsible for execution. To invoke internal transactions, the control of the execution is transferred to the relevant contract, allowing the contract to execute the code and update the state.

% We call contract that an external transaction calls to as \textit{executing contract}, the DApp it belongs to is \textit{executing DApp} and the invoked function is \textit{executing function}. During the execution, there may be internal transactions. Thus, we regard contracts and functions within the same DApp as \textit{associated contracts} and \textit{associated functions}, while calling other DApps, contracts, and functions referred to execution as \textit{non-associated DApps}, \textit{non-associated contracts} and \textit{non-associated functions} respectively.

\begin{figure}[t]
  \begin{subfigure}[b]{0.47\textwidth}
    \begin{lstlisting}[language=Solidity]
contract Pool{
  // The victim function
  function decrease(uint amount, address asset) public nonReentrant  {
    require(allow(msg.sender), "Wrong user!");
    // Calculate balance based on wrong price
    uint256 balance = oracle.getPrice(asset) * amount;
    oracle.doHardWork(msg.sender);
    return balance;
  }
}
contract Oracle{
  function getPrice(address asset) public view {
    require(exist(asset), "Wrong Asset!");
    Vault vault = vaults[asset];
    uint (balance, totalToken) = vault.getFunds(asset);
    return balance / totalToken
  }
  function doHardWork(address account) public {
    if (account == owner) {
        ...
    }
  }
}
    \end{lstlisting}
    \caption{The victim DApp.}
    \label{readOnlyExample1}
  \end{subfigure}
  \begin{subfigure}[b]{0.47\textwidth}
    \begin{lstlisting}[language=Solidity]
contract Vault{
  // The entry function
  function exitVault() public nonReentrant {
    require(allow(msg.sender), "Wrong user!");
    ...
    updateTokens(msg.sender);
    // Control flow transfers to attacker
    transferBalance(msg.sender);
    updateBalances(msg.sender);
    rate = balance / totalToken;
  }
  function swap(address pool) public nonReentrant{
    uint cur_rate = getRate();
    ...
  }
  function setRate(uint newRate) public onlyOwner{
    rate = newRate;
  }
  // The manipulable function
  function getFunds() public view{
    // Return outdated values
    return (balance, totalToken);
  }
  function getRate() public view{
    return rate;
  }
}
    \end{lstlisting}
    \caption{The entry DApp.}
    \label{readOnlyExample2}
  \end{subfigure}
  \caption{An example of Read-Only Reentrancy.}
  \label{readOnlyExample}
\end{figure}

\subsection{Reentrancy and Read-Only Reentrancy}
\label{subsec:read-only}

\para{Reentrancy} In smart contracts, reentrancy is a specific type of vulnerability that exploits unsynchronized updated data~\cite{reentrancySurvey}. Specifically, due to the serialized execution mechanism of smart contracts, there may exist an inconsistency between the global states. For example, when one contract calls another contract, the states of the calling contract do not fully update before control transfer, like not updating the token supply before transferring funds.
% Reentrancy is a type of state synchronization vulnerability~\cite{reentrancySurvey}. Specifically, due to the serialized execution mechanism of smart contracts, there may exist an inconsistency between the global states, when one contract calls another contract and the states of the calling contract not fully update before control transfer, such as not updating the ledger before transferring funds. 
Attackers can exploit this inconsistency by reentering contracts and making a profit. 

In smart contracts, there are three types of functions that are closely related to reentrancy attacks, as explained below: 

\begin{itemize}
    
    \item \textbf{\textit{Entry Function}}. The entry function is the function where the attacker initializes reentrancy. Entry functions are publicly accessible.
    %Attackers need to call the entry function first when launching reentrancy.
    
    \item \textbf{\textit{Victim Function}}. The victim function is the target function of the reentrancy attack. For each reentrancy attack, the victim function suffers the actual damage such as economic loss. 
    
    \item \textbf{\textit{Manipulable Function}}. Manipulable functions are functions that can be controlled by the attacker to achieve a successful reentrancy attack. For example, a function that returns specific token prices as attackers expect.
\end{itemize}

% In this paper, we refer to the location where ROR is initiated as \textit{entry function}, such as the functions that attackers need to call when launching attacks, the compromised location is \textit{victim function}, while the functions that can be manipulated by attackers, like reading expired states, are \textit{manipulable functions}.
% Reentrancy attempts to exploit the execution of internal transactions. In detail, during internal transactions, the state of contracts, such as users' balance, may not be fully updated. In such a scenario, if attackers can hijack the control flow, they can repeatedly call a specific function, such as continuously withdrawing funds. Therefore, if contracts interact with an insecure contract, such as a contract deployed by an attacker or with vulnerabilities, attackers can make a profit. 

\para{Read-Only Reentrancy}
With the increasing complexity of DApp functionalities, such as token swap~\cite{xu2023sok}, lending~\cite{lengdingsok}, and collateral~\cite{defisok}, different DApps may interact with each other for data access~\cite{popescu2020decentralized}. However, such a deep integration and heavy interactions bring new attack surfaces to DApps, raising significant challenges to DApp security. For example, if the state updates between DApps are not well synchronized, it may lead to potential state inconsistencies in different DApps and further cause attacks such as ROR. 
Usually, in traditional reentrancy, the victim function and manipulable function are in the same DApp, while the victim function of ROR is in another DApp. This means that the search space for ROR is broader, making it more challenging to detect.

\begin{figure}[tbp]
    \centering
    \includegraphics[width=0.43\textwidth]{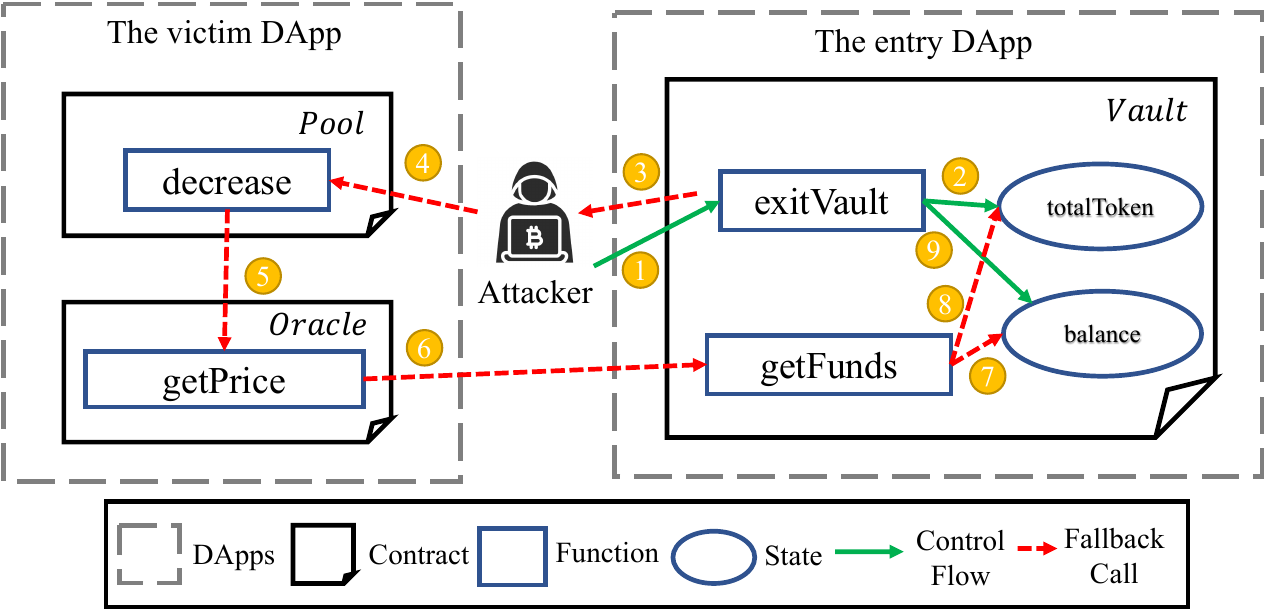}
    \caption{The attack process of example in Fig.~\ref{readOnlyExample}.}
    \label{fig:attackworkflow}
\end{figure}

\begin{figure*}[t]
    \centering
    \includegraphics[width=0.85\textwidth]{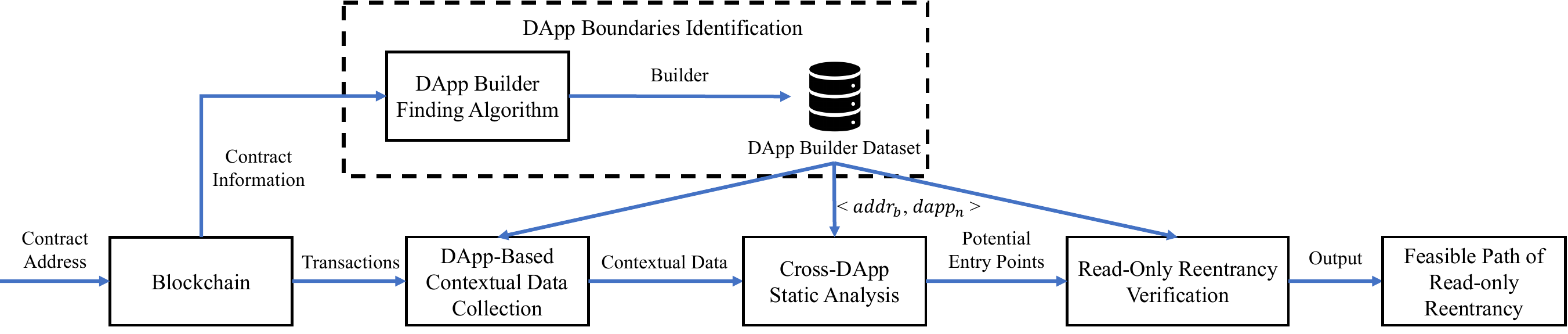}
    \caption{The workflow of \system{}.}
    \label{overview}
\end{figure*}

\subsection{Motivating example}
\label{subsec:motivating_example}
Fig.~\ref{readOnlyExample} shows a simplified code snippet of three smart contracts in two DApps, victim DApp and entry DApp. The victim DApp in Fig.~\ref{readOnlyExample}(\subref{readOnlyExample1}) has a contract \textit{Pool}. It allows users to withdraw balances by executing the victim function \textit{decrease}. Although both DApps have added checks, both line 4 in Fig.~\ref{readOnlyExample}(\subref{readOnlyExample1}) and Fig.~\ref{readOnlyExample}(\subref{readOnlyExample2}), to ensure security, problems arise when they interact with each other. The attack process is shown in Fig.~\ref{fig:attackworkflow}. An attacker can first call the entry function \textit{exitVault} in the entry DApp. During the execution, the control flow of \textit{exitVault} transfers to the attacker while states not fully update (steps 2 and 3 in Fig.~\ref{fig:attackworkflow}), causing a temporary mismatch between token amount and balance. Then, instead of reentering the entry DApp, the attacker executes \textit{decrease} in the victim DApp (step 4 in Fig.~\ref{fig:attackworkflow}) and the price is obtained through the manipulable function \textit{getFunds} of the entry DApp (step 6 in Fig.~\ref{fig:attackworkflow}). However, since state \textit{balance} has not been updated at this point, \textit{decrease} will get an incorrect price. As a result, attackers can get more assets. 

\para{Limitations of existing works} 
Existing methods (both static and dynamic methods) can not effectively identify ROR due to the following unique challenges.

\textit{$\bullet$ Identifying DApp boundaries for contracts in ROR.} Current contracts' DApp information collection methods~\cite{disentanglingDefi, defiIntegration,icychecker} are inaccurate. For example, when analyzing \textit{decrease} in Fig.~\ref{readOnlyExample}(\subref{readOnlyExample1}), contracts and functions under the same DApp, such as contract \textit{Oracle} and function \textit{doHardWork} may be introduced. As a result, the search space becomes large. 

\textit{$\bullet$ Finding entry functions of ROR.} Existing methods for detecting reentrancy~\cite{deeplearningReentrancy, Clairvoyance} cannot trace DApp data, which makes it difficult to identify potential entry functions of ROR. In our example, all functions in contract \textit{Vault} need to be analyzed, causing significant performance overhead and false positives.

\textit{$\bullet$ Verifying the presence of ROR.} Due to the lack of runtime contextual information, existing static analysis techniques such as DeFiTainter~\cite{defiTainter} can hardly detect and verify RORs. In the meantime, state-of-the-art fuzzing approaches~\cite{xFuzz, ityFuzz} have proved effective for vulnerability detection in smart contracts. Unfortunately, these approaches are still inadequate for detecting RORs as their randomly generated fuzzing inputs can hardly trigger RORs with high complexity. For example, it is with little chance to generate valid fuzzing inputs that can bypass line 4 in Fig.~\ref{readOnlyExample}(\subref{readOnlyExample1}) and line 4 in Fig.~\ref{readOnlyExample}(\subref{readOnlyExample2}) simultaneously.

\para{Our solution}
\system{} uses cross-DApp analysis to guide the detection of RORs. Turning to the example in Fig.~\ref{readOnlyExample}, \system{} first collects and replays the transactions of \textit{decrease}. During the replay, \system{} records DApp-based contextual information, such as DApps of called contracts. Then \system{} finds out manipulable functions, like \textit{getFunds}. Such manipulable functions are the key to uncover the entry functions of ROR due to the shared state dependencies with entry functions. For example, since function \textit{exitVault} modifies the state \textit{balance} that \textit{getFunds} relies on, \system{} regards \textit{exitVault} as a potential entry point. \system{} then performs multi-function fuzzing for \textit{exitVault} and \textit{decrease}. In detail, \system{} tries to reenter \textit{decrease} during the execution of \textit{exitVault}. When \system{} finds a reachable path, it reports relevant inputs and functions and explains how an attacker can initialize ROR. Turning to this example, \system{} will report that an attacker can hijack control flow during executing \textit{exitVault} and reenter \textit{decrease}.

\section{Design of \system{}}
\label{chap:design}
% In this section, we first introduce the technical challenges and ideas of \system{} respectively. Then, we present the overview of \system{}.

\subsection{Technical Challenges and Our Idea}
% We summarize the technical challenges and ideas of \system{} as follows:

\para{Identifying DApp boundaries}
Due to the anonymity of the blockchain, smart contracts do not inherently contain information about the DApp it belongs to. To identify DApp boundaries, a straightforward approach is to collect the contract addresses of each DApp from the official websites or third-party APIs (e.g., DAppRadar~\cite{dappRadar} and Thegraph~\cite{theGraph}). However, the DApp information from these sources is mostly outdated and inaccurate, as a large number of new smart contracts are created and deployed every day.

To address this, \system{} identifies the DApp boundaries (i.e., DApp builders) based on creators of various smart contracts. Compared to the contract-DApp mapping information scattered in various sources, information about DApp builders (contract creators) is better documented and easier to access. Therefore, the DApp builder information is a more reliable source to identify the DApp boundary.

\para{Finding potential entry functions}
As mentioned earlier, the entry function and the victim function of ROR have no direct relationship. For example, we can not find a call chain from function \textit{decrease} to function \textit{exitVault}, and vice versa. To find potential entry functions, the simplest approach is to traverse all functions and contracts involved in the execution path. However, this is rather impractical due to the large exploration space, with a significant number of false positives. Existing tools~\cite{smartian, contractFuzzer} can not detect ROR due to the ignorance of these important factors.
%Thus, it is necessary to remove irrelevant functions as much as possible, while retaining potential entry functions.

To narrow down the search space, \system{} employs a two-step filtering process to locate entry functions. The key insight here is that while the entry point of ROR is not directly related to the victim function, it must be associated with manipulable functions in the call chain. For example, in Fig.~\ref{readOnlyExample}, the entry function \textit{exitVault} updates states that the manipulable function \textit{getFunds} depends on. To this end, \system{} first points out all functions of other DApps in the call chain. Then, only functions that share state dependency with those manipulable functions are considered as potential entry functions of ROR.
To achieve this, \system{} collects DApp-based contextual data, including the state changes and invoking of each DApp, and comprehends the entire call chain. These information allows \system{} to further identify all manipulable functions and the candidate ROR entries. 
We will give more details about this process in Section~\ref{subsec:data_collection} and \ref{subsec:cross_dapp_analysis}

% \system{} checks control flow integrity to find out potential entry points. \jw{The reasons are as follows: (1) ROR achieves its malicious intention by hijacking control flow similar to other forms of reentrancy. (2) The entry point of ROR must be associated with functions in the call chain, e.g., functions of other DApps in the call chain.} 

\para{Verifying the existence of ROR}
As mentioned earlier, the last challenge lies in how to provide valid fuzzing inputs that can precisely trigger the critical path of ROR across two DApps. More specifically, with random input generation mechanisms as in prior works~\cite{ityFuzz,sFuzz}, it is hard to satisfy the internal check in both DApps. Even if we bypass the check, there is no guarantee that the input can trigger the ROR path (line 8 in Fig.~\ref{readOnlyExample}(\subref{readOnlyExample2})). 

%As mentioned earlier, the last challenge lies in how to provide effective inputs for multi-function in different DApps poses a non-trivial challenge, as finding data that can precisely trigger the critical path of ROR and flow within multiple DApps is difficult. 

To generate valid inputs for both potential entry functions and victim functions, \system{} utilizes their corresponding on-chain historical transactions to detect ROR. This is because on-chain transaction data, such as the sender, timestamp, or inputs, provides realistic execution environments for fuzzing. Compared with randomly generating such fuzzing inputs as in prior works, the on-chain data allows \system{} to bypass the internal checks, such as \textit{require} statements in different functions. In this way, our fuzzing mechanism can trigger a more complex call chain between the entry function and the victim function.
Besides, the successful execution of the combined transactions indicates the existence of the vulnerable path of ROR. More details of this fuzzing process are elaborated in Section~\ref{subsec:verification}.

%\jw{as on-chain data can help us pass function internal checks and trigger deeper paths.} It is feasible to simulate ROR attacks by combining two transactions because even though two transactions' execution environments, such as the sender, timestamp, or block hash, may not be the same, the successful execution of the combined transactions indicates the existence of the vulnerable path. In this way, an attacker can launch ROR by only depositing tokens and invoking corresponding functions in both DApps. 

%We will show how \system{} performs it in Section~\ref{subsec:verification}.

\begin{figure}[t]
    \centering
    \begin{subfigure}{0.25\textwidth}
        \centering
        \includegraphics[width=\textwidth]{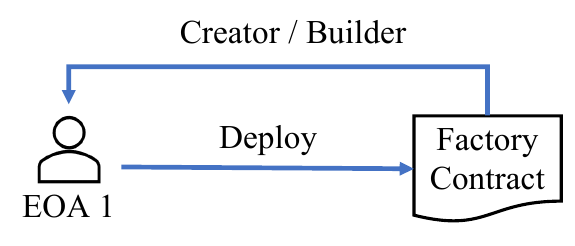}
        \caption{Direct Deployment.}
        \label{createSmartContract_direct}
    \end{subfigure}
    \hfill
    \begin{subfigure}{0.44\textwidth}
        \centering
        \includegraphics[width=\textwidth]{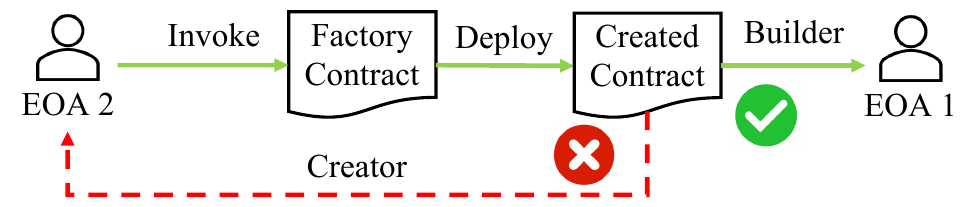}
        \caption{Indirect Deployment.}
        \label{createSmartContract_indirect}
    \end{subfigure}
    
    \caption{Two methods for deploying smart contracts.}
    \label{createSmartContract}
\end{figure}

\subsection{Workflow of \system{}}
% Fig.~\ref{overview} shows the workflow of \system{}. It mainly consists of four modules, (1) DApp Boundaries Identification, (2) DApp-based Contextual Data Collection, (3) Cross-DApp Control-flow Integrity Analysis, and (4) Read-Only Reentrancy Verification. 

Fig.~\ref{overview} shows the workflow of \system{}. Specifically, given an unknown contract address as input, \system{} outputs its corresponding ROR detection results along with the valid ROR paths if exist.

In DApp Boundaries Identification, to identify which DApp the contract belongs to, \system{} retrieves a set of information related to contract creation from the blockchain, including the deployment transaction, internal transaction list of deployment transaction, and the transaction sender. The above information helps \system{} to get the actual builder of the contract, and further identify the DApp it actually belongs to. We will detail this process in Section~\ref{subsec:DApp_identification}. 
%\jw{In the subsequent analysis, these data will ensure that \system{} can accurately identify the boundaries between DApps.}

In DApp-based Contextual Data Collection, \system{} collects the contract's historical transactions, faithfully replays them, and filters DApp-based contextual data by DApp boundaries.
In Cross-DApp Static Analysis, \system{} uses contextual data to extract potential entry points (entry functions) of RORs.
In Read-Only Reentrancy Verification, \system{} employs a multi-function fuzzing strategy to verify the existence of ROR in potential entry points and output feasible ROR paths if exist.

% To identify which DApp a contract belongs to, \system{} employs DApp Identification method to construct a DApp builder dataset $D_{builder}$ and uses $D_{builder}$ to guide the subsequent process.
% %
% To obtain DApp-based contextual data, \system{} replays historical transactions of contracts, and collects fine-grained contextual data during execution with the help of DApp information.
% %
% Then, \system{} tries to find potential entry functions, like \textit{exitVault}, with cross-DApp control-flow integrity analysis. 
% %
% Finally, \system{} tries to hijack the control flow with a multi-function fuzzing strategy to verify the existence of ROR.

\section{Approach Details}
\label{chapt:approach_details}

\begin{algorithm}[t]
        \footnotesize
        \caption{DApp Builder Finding Algorithm.}
        \label{algor:dapp_builder_finding}
        % \scriptsize
        \renewcommand{\algorithmicrequire}{\textbf{Input:}}
        \renewcommand{\algorithmicensure}{\textbf{Output:}}
        
        \begin{algorithmic}[1]
            \REQUIRE Contract address $addr_{s}$  %%input
            \ENSURE Contract builder $builder$    %%output
            
            \STATE  $Tx_{D}$ := \textit{fetchDeploymenTx}($addr_{s}$);
            \STATE  $creator$ := \textit{fetchCreator}($addr_{s}$);
            \STATE  $L_{D}$ := \textit{fetchInternalTxList}($Tx_{D}$);
            \WHILE{\textit{isNotEmpty}($L_{D}$) \&\& \textit{contain}($L_{D}$, $addr_{s}$)}
                \STATE $addr_{in}$ := \textit{fetchFactoryContract}($addr_{s}$);
                \STATE  $Tx_{D}$ := \textit{fetchDeploymenTx}($addr_{in}$);
                \STATE  $creator$ := \textit{fetchCreator}($addr_{in}$);
            \STATE  $L_{D}$ := \textit{fetchInternalTxList}($Tx_{D}$);
            \ENDWHILE  
            \STATE $builder$ := $creator$;
            \RETURN $builder$
        \end{algorithmic}
\end{algorithm}

% \begin{algorithm}[t]
% \footnotesize
% \caption{Enhancing the Robustness of LLM Online Code Watermark via Code Structure and Syntax}
% \label{Alg:Alg2}
% \textbf{Input}: Existing tokens: $R^{(l)}=\left\{t_0, t_1, \cdots , t_m, v_i^{\prime}, v_i^{\prime \prime}, \cdots, v_i^{(l)}\right\}$ , watermark to be embedded $w_X$, set $\widehat{A,B,C,D}$.   
% \begin{algorithmic}[1]
% \IF{$l \leq L$}
% \STATE Check $v_i^{(l)}$ only whitespace and rollback or keep the $X$. $\# \textbf{Pattern 5}$
% \IF{not $\mathcal{LOCK}$}   % #not $\mathcal{LOCK}
% \IF{$v_i^{(l)} \in \left\{ \widehat{A} \cup \widehat{B} \cup \widehat{C} \cup \widehat{D} \right\}$}
% \STATE $ \mathcal{LOCK} \Leftarrow 1$.  \quad \quad  \quad  \quad   \quad  \quad \quad  \quad $\# \textbf{Pattern 6}$
% \STATE Rollback and skipping watermark information based on different patterns triggered by $v_i^{(l)}$ and update $w_X$. $\# \textbf{Pattern 1, 2, 3, 4}$
% \STATE $V = V$, \textbf{break}.
% \ELSE  % #not $\mathcal{LOCK}
% \IF{$X \leq x$} 
% \STATE Take set $D$ or $V \cap D$ corresponding to $w_X$, to $V$, $X = X+1$.
% \ELSE
% \STATE X = 0, Iterative embedding the watermark. \quad $\# \textbf{Pattern 7}$
% \ENDIF
% \ENDIF
% \ENDIF
% \STATE Based on the effectiveness of Pattern to determine $ \mathcal{LOCK} \Leftarrow 0$.
% \STATE $V = V$, \textbf{break}.
% \ENDIF
% \end{algorithmic}
% \end{algorithm}

\subsection{DApp Boundaries Identification}
\label{subsec:DApp_identification}

\system{} uses DApp builders to identify DApp boundaries. Note that due to the unique DApp development mode, the builder of a given DApp cannot simply attributed to the creator(s) of the smart contracts it contains. 
As the examples shown in Fig.~\ref{createSmartContract}, there are two ways to deploy smart contracts, namely \textit{direct deployment} and \textit{indirect deployment}. For direct deployment (Fig.~\ref{createSmartContract}(\subref{createSmartContract_direct})), users can deploy a new contract by directly sending a transaction to the blockchain. In this case, the builder of contract \textit{Factory} is its creator \texttt{EOA1}.
Differently, a more complicated case is the scenario of \textit{indirect deployment}. As shown in Fig.~\ref{createSmartContract}(\subref{createSmartContract_indirect}), users can send a transaction to a contract (i.e., \textit{Factory Contract}) that further creates another new contract (\textit{Created Contract}). Since in smart contract, a contract's creator is always recorded as the account that invokes the transaction, simply using this information from transaction records can easily cause misclassification. In this case, the actual builder of \textit{Created Contract} should be \textit{EOA1}, not its creator \textit{EOA2}. 

%Here, the builders of \textit{Factory Contract} and \textit{Created Contract} are \texttt{EOA1} and \texttt{EOA2} respectively. However, in reality, \textit{Created Contract} is considered as \texttt{EOA1}'s contract, because its functionality is determined by \textit{Factory Contract}, which \texttt{EOA1} creates. 

\para{DApp builder identification} To identify the builder of a given contract, \system{} employs a heuristic-based DApp Builder Finding (DBF) algorithm, as shown in Algorithm~\ref{algor:dapp_builder_finding}. The core idea here is to exhaustively find the actual builder of the contract, based on various information related to contract creation.
In detail, to find the builder $builder$ of a contract $addr_{s}$, \system{} first fetches its deployment transaction $Tx_D$, the first transaction of the contract, from the blockchain and searches for the creator $creator$ in $Tx_D$. Then \system{} analyzes how $Tx_D$ is deployed by checking the internal transaction list $L_D$, which records all internal transactions executing in $Tx_D$. When $addr_{s}$ is created by indirect deployment, there will be an internal creation transaction related to $addr_{s}$ in $L_D$. Thus, if $L_D$ is empty or \system{} can not find $addr_s$ in $L_D$, it means $addr_{s}$ is created by direct deployment and $builder$ is $creator$. Otherwise, \system{} gets the internal creator $addr_{in}$, which means $addr_s$ is deployed by another smart contract $addr_{in}$ and they belong to the same DApp. Then, \system{} checks $Tx_D$, $creator$, and $L_D$ of $addr_{in}$ iteratively until $L_D$ is empty or $addr_{in}$ not exists in $L_D$. In this way, \system{} finally obtains $builder$ of $addr_{s}$. 

We use the examples in Fig.~\ref{createSmartContract} to illustrate this process. For \textit{Factory Contract} in Fig.~\ref{createSmartContract}(\subref{createSmartContract_direct}), \system{} can find its $creator$ is \texttt{EOA1}. As \textit{Factory Contract} is directly deployed, its $L_D$ is empty. Thus, \system{} can verify that \texttt{EOA1} is the builder of \textit{Factory Contract}. For \textit{Created Contract} in Fig.~\ref{createSmartContract}(\subref{createSmartContract_indirect}), its $creator$ is \texttt{EOA2}. As \textit{Created Contract} is deployed indirectly, its $L_D$ is not empty and \system{} can find the internal builder $addr_{in}$ of \textit{Created Contract} is \textit{Factory Contract}. In this case, \system{} considers both of the two contracts to belong to the same DApp. Therefore, \system{} investigates how \textit{Factory Contract} is deployed. Since \textit{Factory Contract} is directly deployed by \texttt{EOA1}, \system{} can determine that the builder of the \textit{Created Contract} is \texttt{EOA1} instead of \texttt{EOA2}.

% We calrified that our approach in identifing DApp boundaries may not be sound. However, this does not impact the effectiveness of SmartReco for detecting ROR, as we will treat all unknown contracts as unsafe ones, which will further undergo subsequent security checks.

Note that, \system{} may fail to identify some contracts if their DApp builders are totally absent from $D_{builder}$. This could potentially lead to inaccuracies in boundary identification. However, this will not diminish the effectiveness of \system{} in detecting ROR. Specifically, \system{} employs a conservative approach to address this potential misinformation. In \system{}, all contracts whose builders are not listed in $D_{builder}$ will be deemed unsafe, and these contracts will then undergo further security analysis.

% \yuhong{
% \para{Enhancing identification robustness} Note that, since \system{} only relies on a limited set of labeled contracts to build $D_{builder}$, \system{} may miss builders of some DApps. To ensure the accuracy of DApp boundaries identification, for contracts whose builders are not in $D_{builder}$, \system{} only considers the interactions between contracts from the same builder as secure, while all others will be considered unsafe and conduct a strict inspection. We will demonstrate in Section~~\ref{subsec:effeciveness_of_each_module} that using builders to identify DApp boundaries is effective. 
% }

\subsection{DApp-based Contextual Data Collection}
\label{subsec:data_collection}
In this step, \system{} replays transactions as in prior works~\cite{icychecker,offchain,Time-travel} to obtain DApp-based contextual information such as contract address and state read and write. The difference is that \system{} replays at transaction level, while others perform based on synchronizing blockchain state. The advantage is that \system{} does not require a significant amount of time and storage to gain the entire blockchain state and can be easily extended to other blockchain platforms like BNB Chain~\cite{bnb}. More specifically, during the replay of a transaction, whenever \system{} needs to invoke another contract $addr_s$, it will first identify the DApp information $dapp_n$ of this contract and record the $operation$ executed in $addr_s$ in a three-tuple \textless\textless$addr_{s}$, $dapp_{n}$\textgreater, $operation$\textgreater. For example, when \system{} meets opcodes related to state read and write, such as \textit{SLOAD} and \textit{SSTORE}, \system{} records \textit{read} and \textit{write} in $operation$ respectively. Similarly, \system{} records invocation information whenever it encounters opcodes related to method execution, such as \textit{CALL}, \textit{CALLCODE}, \textit{DELEGATECALL}, and \textit{STATICCALL}. The $operation$ here is \textit{invoke}. We will demonstrate that replay at transaction level is feasible in Section~\ref{subsec:effeciveness_of_each_module}. 

\subsection{Cross-DApp Static Analysis}
\label{subsec:cross_dapp_analysis}
% Potential Entry Functions Determination \mx{generation-based fuzzing or Control-flow Integrity Manipulation}
% \mx{
% consider two ways of storytelling:

% (1) A generation-based fuzzing module. We first do seed scheduling with previously initialized dataset. Then we use four strageties to distill other seed and to generate new seeds.

% (2) we try to break control-flow integrity here. We first find malicious function for injection. Then we find potential entry point and hijack the corresponding control-flow. Note that, we first find malicious function then we find potential entry point due to the ROR characteristics (Previous control-flow hijacking work did it in a reversal way).
% }
This step comprises two tasks: (1) prioritizing manipulable functions from a given call chain for efficiency, and (2) identifying the set of potential entry functions in other DApps based on the selected manipulable functions.

%\system{} uses DApp-based contextual data to locate manipulable functions in transactions and \jw{uses cross-DApp static analysis to} identify potential entry functions.

\para{Manipulable function prioritization}
Given the huge amount of manipulable functions in cross-DApp interactions, \system{} needs to prioritize their order as it can significantly improve detection efficiency. 
Thus, \system{} introduces a metric called \textit{importance} to order the manipulable functions. Here, \textit{importance} is calculated based on the sum of contextual data (i.e., invoke, read, and write) of each function in the call chain. Our main idea is that if a function frequently appears or performs many operations in historical transactions replay, it is more likely to provide high-quality contextual information (i.e., state dependency) that is critical to find ROR entry functions. The calculation formula is as follows:
\begin{align}
importance = C_{invoke} + C_{read} + C_{write}
\end{align}
where $C$ represents the count.

To this end, \system{} generates the inter-DApp data flow graph based on DApp-based contextual data. The graph records the operations in each manipulable function, such as \textit{call}, \textit{read}, and \textit{write}. Then, \system{} orders these manipulable functions based on their \textit{importance} in descending order.

\begin{figure}[tbp]
    \centering
    \includegraphics[width=0.47\textwidth]{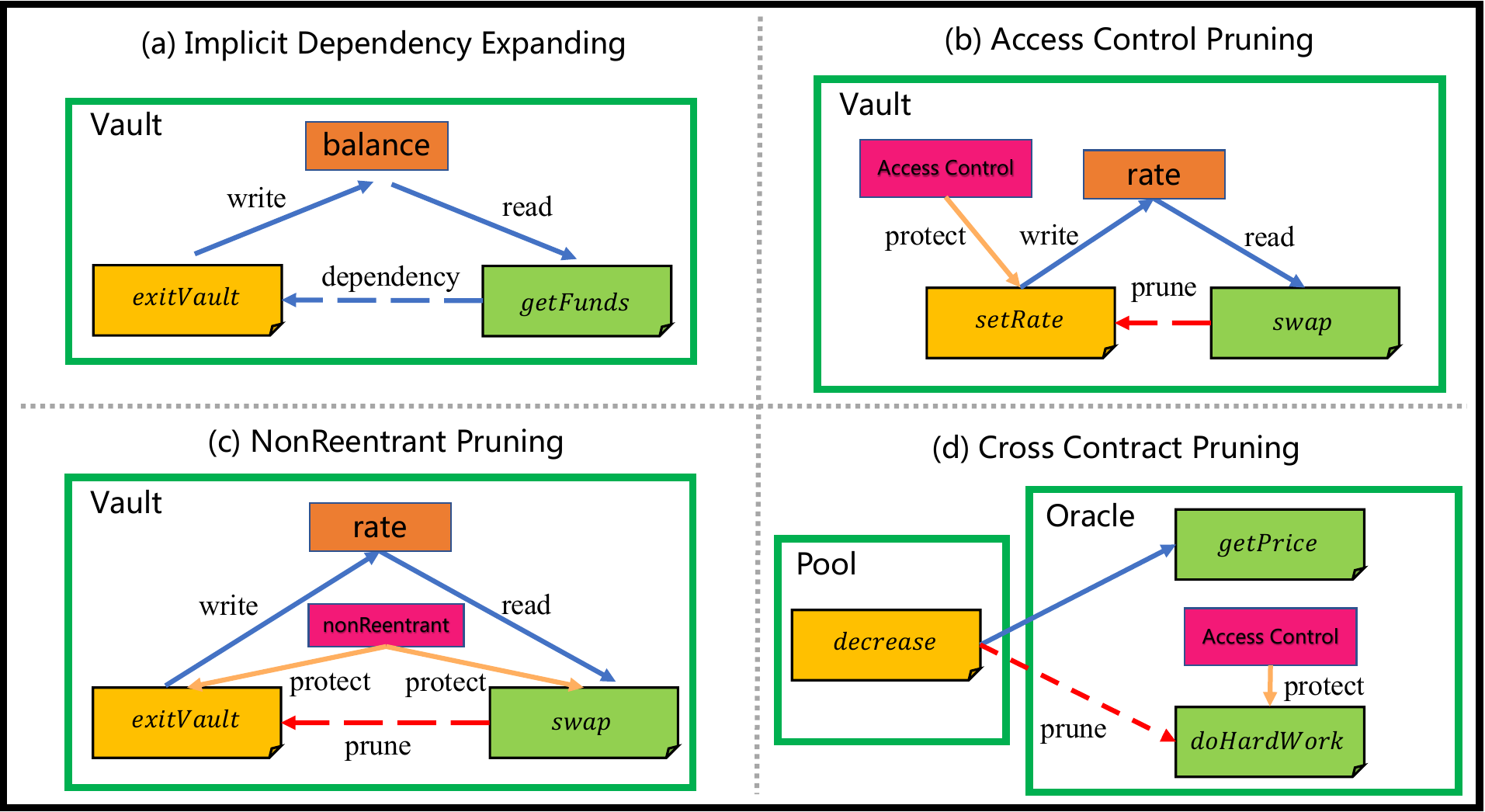}
    \caption{Rules for constructing intra-DApp graph.}
    \label{interdappRules}
\end{figure}

\para{Potential entry functions determination}
Based on the characteristics of real-world ROR attacks, we have summarized four rules (shown in Fig.~\ref{interdappRules}) for constructing an intra-DApp graph and identifying potential entry functions. To the best of our knowledge, these rules are relatively comprehensive, as they sufficiently cover all known ROR attacks (8 in total). Meanwhile, it is practical to incorporate new rules into \system{} should new varieties of ROR attacks emerge. Below, we use contracts in Fig.~\ref{readOnlyExample} to show how \system{} constructs the intra-DApp graph.
\begin{itemize}
    \item \textbf{Implicit dependency expanding.} When there are two functions in a contract, where one modifies the state and the other reads the state, \system{} regards these two functions have an implicit dependency and adds an edge between them. For example, in Fig.~\ref{interdappRules}(a), \textit{exitVault} modifies the state \textit{balance}, while \textit{getFunds} reads \textit{balance}. \system{} will add an implicit dependency from \textit{getFunds} to \textit{exitVault}.
    
    \item \textbf{Access control pruning.} Based on access control, \system{} performs implicit dependency pruning. More specifically, if the modification of states within functions is protected by access control, like \textit{setRate} in Fig.~\ref{interdappRules}(b), \system{} does not consider it as an unsafe function and prunes all implicit dependency edges that depend on it, like \textit{swap} in the example.
    
    \item \textbf{NonReentrant pruning.} Although modifier \textit{nonReentrant} is ineffective in defending against ROR. However, within the same contract, \textit{nonReentrant} is feasible. Specifically, when two functions in the same contract are both protected by \textit{nonReentrant}, like \textit{exitVault} and \textit{swap} in Fig.~\ref{interdappRules}(c), even if they have a relationship on the same state, they are unable to cause ROR. Therefore, it is safe to remove the dependency edge between them.

    \item \textbf{Cross contract pruning.} As mentioned before, ROR is a cross-DApp attack. For functions in the same DApp with sufficient access control, \system{} considers them as safe functions and prunes their corresponding execution paths. As shown in Fig.~\ref{interdappRules}(d), since \textit{doHardWork} in contract \textit{Oracle} is protected by access control, line 19 in Fig.~\ref{readOnlyExample} (\subref{readOnlyExample1}), \system{} prunes \textit{doHardWork} in \textit{decrease}.
\end{itemize}

Based on the above rules, \system{} can construct an intra-DApp control flow graph. More specifically, for any manipulable function, all endpoints of implicit dependency edges have the potential to become an entry point and need further testing, like \textit{exitVault} in Fig.~\ref{readOnlyExample}(\subref{readOnlyExample2}).

\subsection{Read-Only Reentrancy Verification}
\label{subsec:verification}
To verify ROR, \system{} generates valid inputs for entry functions and attempts to trigger ROR by performing control flow hijacking.
%, as ROR, like other reentrancies~\cite{icychecker}, attacks by breaking the control flow. }

\para{Inputs generation}
For each potential entry function $fun$, \system{} attempts to generate its valid input during the replay of a transaction $tx_o$ extracted from the detected contract. In detail, if there is no historical transaction of $fun$, \system{} will use the same environment in $tx_o$, such as the same caller, to construct a transaction of $fun$. More specifically, \system{} will generate transaction $tx_{fun}$ based on the ABI of $fun$ and add it to candidate list $l_c$. Otherwise, if historical transaction $tx_{fun}$ of $fun$ is fetched from the blockchain like Ethereum~\cite{ethereum}, \system{} pushes it into the $l_{c}$. Since most normal transactions may not trigger the critical path of ROR, \system{} mutates $tx_{fun}$ using the following two strategies:
\begin{itemize}
    \item \textbf{Fuzzing funds}. If $fun$ is \textit{payable}, then it can receive funds, like ETH, and should have complete logic to handle the assets involved in the transaction. Therefore, \system{} randomly changes funds values in $tx_{fun}$, gets a new transaction, and adds it to $l_{c}$.
    \item \textbf{Fuzzing input.} If $fun$ can accept parameters, to explore unusual paths, \system{} will attempt to mutate them randomly based on the original inputs in $tx_{fun}$, get a new transaction, and add it to $l_c$.
\end{itemize}
% \begin{figure}[tbp]
%     \centering
%     \begin{subfigure}{0.22\textwidth}
%         \centering
%         \includegraphics[width=\textwidth]{picture/hjack_origin_process.pdf}
%         \caption{execution path of $tx$.}
%         \label{hjack_demo_origin}
%     \end{subfigure}
%     \hfill
%     \begin{subfigure}{0.23\textwidth}
%         \centering
%         \includegraphics[width=\textwidth]{picture/hjack_hajck_process.pdf}
%         \caption{execution path of $(tx,tx_{o})$.}
%         \label{hjack_demo_hjack}
%     \end{subfigure}
    
%     \caption{Potential entry function verifying process.}
%     \label{hjack_demo}
% \end{figure}

\begin{figure}[tbp]
    \centering
    \includegraphics[width=0.49\textwidth]{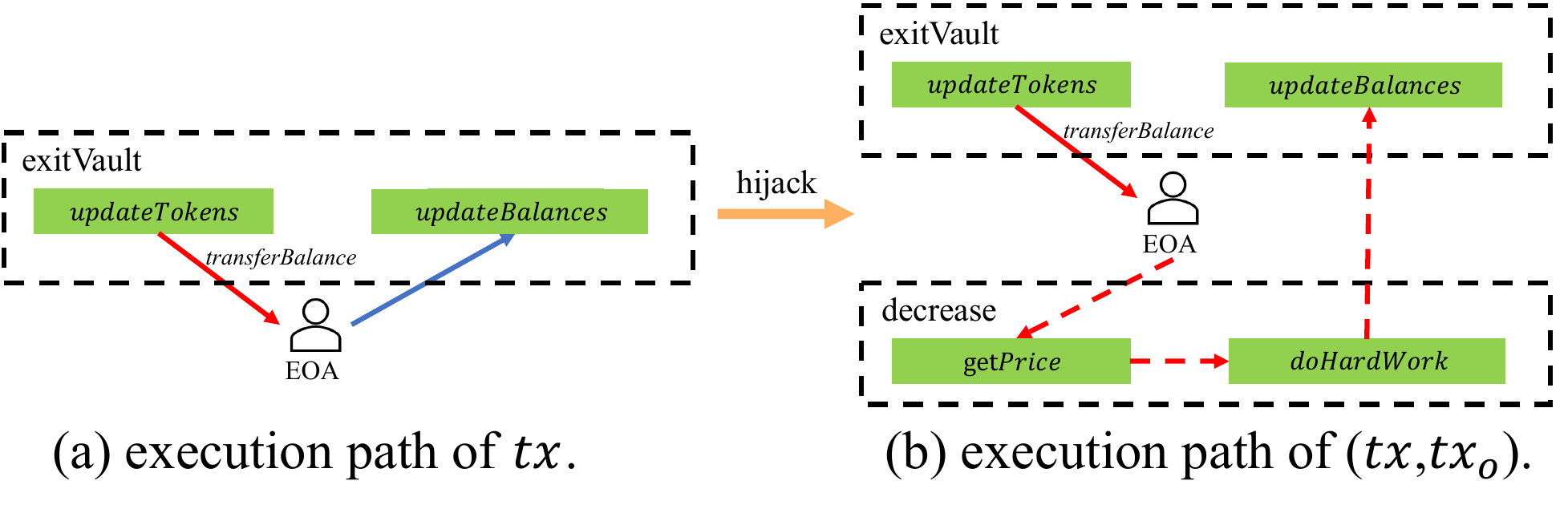}
    \caption{The process of verifying potential entry function for contracts in Fig.~\ref{readOnlyExample}.}
    \label{hjack_demo}
\end{figure}

\para{ROR verification}
For each transaction $tx$ in $l_{c}$, \system{} traces the execution of $tx$ and attempts to hijack its control flow. In detail, whenever $tx$ interacts with an external parameter (e.g., an arbitrary contract address), \system{} tries to invoke $tx_o$. If $tx_o$ executes successfully, \system{} continues to track the execution of $tx$ and monitors state modifications, e.g., updating balances. If $tx$ updates states that $tx_o$ has read, \system{} considers the presence of ROR and reports it. %Otherwise, \system{} will conduct the next round of testing.

We use contracts in Fig.~\ref{readOnlyExample} as an example. During replaying the transaction $tx_o$ of \textit{decrease}, \system{} identifies \textit{exitVault} as a potential entry function. \system{} then generates the transaction $tx_{fun}$ for \textit{exitVault}, mutates $tx_{fun}$ to create candidate list $l_c$ and analyzes each transaction $tx$ in $l_c$. Fig.
~\ref{hjack_demo}(a) shows the original execution path of $tx$. Specifically, when $tx$ executes to line 8 in Fig.~\ref{readOnlyExample}(\subref{readOnlyExample2}), the execution control is handed over to \texttt{EOA} to perform the transfer. \system{} then simulates an attacker hijacking the control flow and executes $tx_o$, as shown in Fig.
~\ref{hjack_demo}(b). Since $tx_o$ executes successfully and $tx$ modifies the state \textit{balance} that $tx_o$ depends on after executing $tx_o$, \system{} will report the presence of ROR.

\section{Evaluation}
\label{chap:evaluation}
In this section, we first present the dataset used in the evaluation and introduce our experimental setup of \system{}. Then we show the evaluation results of \system{}. %Finally, we discuss the results of \system{}.
\subsection{Implementation and Evaluation Setup}
\label{subsec:Implementation}

\noindent{\bfseries Dataset.} We construct the following three datasets to perform our evaluation.

$\bullet$ \textbf {Manual-labeled ROR Dataset ($D_{labeled}$)}. This dataset contains 45 ROR vulnerabilities from 25 contracts through a carefully designed manual-labeling process. Particularly, 14 RORs from 8 contracts are collected from publicly reported attack incidents, while the remaining 31 RORs from 17 contracts all written in Solidity~\cite{dannen2017introducing} are manually discovered by our security experts. More details are shown in Section~\ref{subsec:threats}.
%\jw{Besides, we avoid all contracts written in non-Solidity languages, e.g., Vyper~\cite{vyper}, during the manual data screening process. More details are shown in Section~\ref{subsec:threats}.}} 
To the best of our knowledge, this is the most comprehensive and up-to-date ROR dataset. 

The manual labeling process is performed by three researchers, including one professor and two senior PhD candidates. Each researcher has 2+ years of experience in auditing smart contract vulnerabilities. To ensure the quality of the labeled dataset, we first give a detailed tutorial about ROR to researchers, helping them better understand the nature of RORs. 
Then, we collect all the 8 publicly reported ROR attacks, and extract their entry functions and manipulable functions. We then ask researchers to independently extract the attack patterns (i.e., call chains) from these RORs. Meanwhile, to construct a potential ROR set, we use a similarity-matching approach to search for functions that call manipulable functions in all historical transactions. Finally, we collect 67 suspicious functions from 36 contracts. Based on the extracted ROR patterns, we attempt to construct call chains from the entry functions to functions in the set with the multi-function fuzzing method. Subsequently, the experts independently verify these cases and align disagreements together.
Note that during the whole labeling process, only the vulnerabilities confirmed by all three researchers are considered as RORs. While this criterion is relatively conservative, our labeled dataset provides a lower bound for evaluating the effectiveness of \system{}, excluding potential FPs caused by data labeling.

$\bullet$ \textbf{ Popular DApp Dataset ($D_{unknown}$)}. % To build the DApp builder dataset and validate the effectiveness of the \system{}, 
To further understand the impact of ROR in real-world DApps, we construct an unknown dataset ($D_{unknown}$) consisting of 2,676 smart contracts from 123 popular DApps. More specifically, this dataset is a union of public-available, top-popular DApp data released by two prior research (Stefan et al.~\cite{disentanglingDefi} and IcyChecker~\cite{icychecker}). The identifications of 2,676 smart contracts are determined based on the identified DApp boundaries by \system{}.
%Then, we refine the dataset by employing the DBF algorithm (in Section XX), as some smart contracts are misclassified by prior research. We supplement certain DApp information to ensure the accuracy of the analysis. 
%Finally, $D_{unknown}$ contains 2,676 contracts from 123 DApps. 
%Note that, due to the significant time overhead of fuzz testing, we do not include a larger number of DApps. 
Since all contracts we analyzed are from popular DApps, we believe the analysis results are sufficiently representative of the smart contract ecosystem.

$\bullet$ \textbf{ DApp builder dataset ($D_{builder}$)}. We use all DApps in $D_{unknown}$ to construct the builder dataset with the DBF algorithm described in Section~\ref{subsec:DApp_identification}. All data in $D_{builder}$ are recorded in a tuple \textless$builder$, $dapp_{n}$\textgreater, consisting of each unique builder address and DApp(s) it belongs to. 
Note that for DApps belonging to the same project, if these DApps share the same builder, we will merge these DApps and their builder as a single item in $D_{builder}$, as they are mutually trusted. 
Finally, our $D_{builder}$ consists of a total number of 334 builders and their corresponding DApps. To this end, given the builder of an unknown contact, $D_{builder}$ can tell which DApp the contract exactly belongs to.
%To use $D_{builder}$ to identify DApp boundaries, one needs to provide the builder addresses, either through the DBF algorithm or by manually providing them.

%To construct $D_{builder}$, \system{} analyzes each DApp in $D_{unknown}$ by adopting the DBF algorithm. 

%Besides, for DApps belonging to the same project, there are two strategies to manage them. One is to deploy different DApps using different builders. The other is to use the same deployers for all DApps. To ensure unambiguous results, whenever a new builder is added to the dataset, \system{} checks whether it already exists in other DApps. If so, \system{} considers these DApps to belong to the same project and merge them into one. Otherwise, even if they belong to the same project, \system{} will not merge them and perform more thorough inspections.

%\system{} iterates using the DApp builder finding algorithm to analyze $D_{unknown}$. In detail, for a contract of DApp $dapp_{n}$, if the contract builder $builder$ is not found in $D_{builder}$, \system{} will add $builder$ to the dataset in a tuple \textless$builder$, $dapp_{n}$\textgreater. 
%

\para{Implementation}
We implement \system{} with around 1,700 lines of code in Python and about 2,300 lines of code in Rust. \system{} replays historical transactions and verifies RORs based on ityFuzz~\cite{ityFuzz}, which is an online fuzzing-based framework for smart contracts. \system{} finds potential entry functions based on Slither~\cite{slither}. All experiments in our evaluation are conducted on a machine with two Intel(R) Xeon(R) Gold 5218R CPU @ 2.10GHz, 512GB RAM, and Ubuntu 20.04.4 OS.

\begin{table}[tbp]
\centering
\caption{Overall effectiveness of \system{} on $D_{labeled}$.}
\label{tab:recall_rate}
\small
\resizebox{0.48\textwidth}{!}{

\begin{tabular}{c|ccc|ccc}
\hline
\multirow{2}{*}{$D_{label}$} & \multicolumn{3}{c|}{Recall} & \multicolumn{3}{c}{Precision} \\ \cline{2-7} 
                           & TP     & FN    & Rate       & TP      & FP      & Rate      \\ \hline
Attack incidents           & 8      & 6     & 57.14\%    & 8       & 0       & 100\%      \\ \hline
Manual annotation          & 31     & 0     & 100\%      & 31      & 5      & 86.11\%     \\ \hline
Total                      & 39     & 6     & 86.67\%    & 39      & 5       & 88.64\%      \\ \hline
\end{tabular}
}

\end{table}
\para{Evaluation setup}
We inspect the latest 1000 transactions for each contract and use the top 300 transactions of each potential entry function to test and manually analyze the reported results. If \system{} needs data during execution, such as reading storage, it directly fetches the corresponding storage from the blockchain and caches it locally. Besides, like other transaction-based frameworks (e.g., sFuzz~\cite{sFuzz} and Smartian~\cite{smartian}), \system{} implements a customized Ethereum Virtual Machine (EVM)~\cite{ethereum} to execute transactions, monitor the opcodes, and record data.

\para{Evaluation Metrics}
Specifically, we focus on the following research questions.
\begin{enumerate}[label=\textbf{RQ\arabic*.},leftmargin=0.95cm]
    \item How effective is \system{} in detecting ROR?
    \item What is the impact of each module in \system{} on detecting ROR?
    \item Is \system{} more effective in detecting ROR compared to other advanced tools?
    \item Can \system{} effectively detect unknown ROR in real-world smart contracts?
    \item What is the performance overhead of \system{}  for detecting ROR?
    
    % \item Can DApp identification method accurately identify which DApp a particular contract belongs to?
\end{enumerate}

\subsection{Effectiveness of \system{}}
\label{subsec:SmartReco}
To answer RQ1, we use $D_{labeled}$ to test \system{}, and the results are shown in Table~\ref{tab:recall_rate}. In detail, \system{} successfully detects 39 out of 45 RORs with a precision of 88.64\% and a recall of 86.67\%. In addition, thanks to the fine-grained cross-DApp analysis, \system{} successfully identifies 4 attack contracts out of the 8 attack incidents and accurately pinpoints the specific attack paths.

\para{False Negatives}
Five out of six false negatives are all written in Vyper~\cite{vyper}, which is outside of our scope. Due to \system{} relying on Slither, \system{} cannot detect such contracts. In fact, there are currently no stable static analysis tools specifically for Vyper~\cite{defiTainter}. If the state dependencies of these contracts are provided, \system{} can determine the dependency relationship between functions and identify potential entry functions for analysis. Another false negative is due to the use of a modifier that has similar logic to \textit{onlyOwner}, causing \system{} to mistakenly assume that the function does not need to be analyzed.

\para{False Positives}
After analyzing the false positive cases, we identify the main reasons as follows: (1) When searching potential entry functions, \system{} primarily focuses on checking modifiers because most smart contracts use modifiers for control. However, some contracts do not follow these patterns and instead perform checks within the function, resulting in \system{} failing to recognize them and causing false positives. (2) Some DApps have forwarding modules, such as the Gnosis multi-signature wallet. It helps users to forward and execute operations directly. However, \system{} currently considers only interactions in the same DApp as safe and thus leads to false positives.

\begin{tcolorbox}[colback=SeaGreen!10!CornflowerBlue!10,colframe=RoyalPurple!55!Aquamarine!100!, title = {Answer to RQ 1:}]
% \small
\system{} demonstrates its effectiveness in ROR detection with a precision of 88.64\% and a recall of 86.67\% on the manually-labeled dataset $D_{labeled}$.
\end{tcolorbox}

% While our DApp dataset contains 123 DApps, it is important to acknowledge that there are hundreds and thousands of DApps. Therefore, it is inevitable that we may miss information about certain DApps. When \system{} is unable to determine which DApp a contract should belong to, it marks the contract as "unknown" and considers it potentially unsafe. In such cases, a strict inspection strategy is employed to ensure safety. In this way, the approach may lead to misclassification. However, once we complete the missing DApp information, this false positive is eliminated. 

\begin{figure}[tbp]
    \centering
    \includegraphics[width=0.4\textwidth]{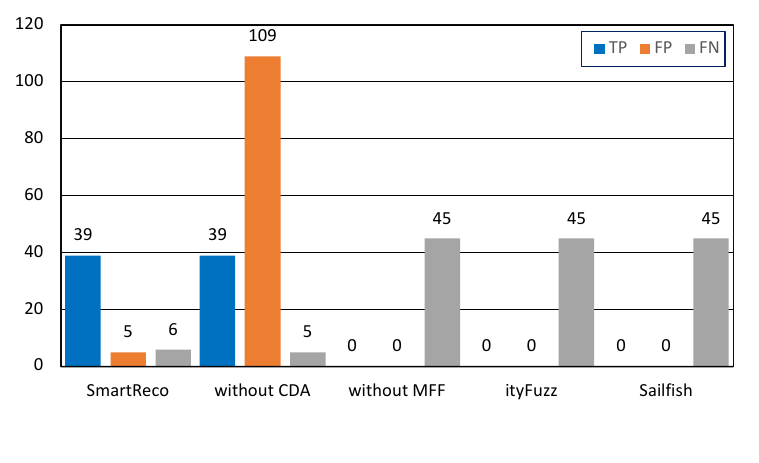}
    \caption{Result of \system{}, \system{} without cross-DApp analysis (without CDA), \system{} without multi-function fuzzing (without MFF), ityFuzz and Sailfish based on $D_{labeled}$. }
    \label{fig:compare_result}
\end{figure}

\subsection{Effectiveness of Each Module in \system{}}
\label{subsec:effeciveness_of_each_module}
To answer RQ2, we evaluate the effectiveness of DApp boundaries identification method with top-10 DApps in $D_{unknown}$. Besides, we verify the effectiveness of the transaction-based replay method with $D_{unknown}$. We also conduct ablation experiments on cross-DApp analysis and multi-function fuzzing methods based on $D_{labeled}$ separately.

\para{Effectiveness of DApp boundaries identification}
To verify whether the newly created contracts will cause the DApp boundaries identification method to fail, we select the top-10 DApps from $D_{unknown}$, as top DApps provide more timely and comprehensive maintenance of contract information. Specifically, we collect the latest contracts of all these DApps from their official websites, totaling 545 contracts. Then, we use the DBF algorithm to find the builders of these contracts, determine whether these builders are in the $D_{builder}$, and whether the corresponding builders are classified correctly. The results show that the DApp boundary identification method accurately recognizes 489 contracts, reaching a precision of 89.72\%. The remaining 56 contracts are due to the contract builders not being included in $D_{builder}$. Overall, our method does not produce any misclassification.
% As shown in Table~\ref{tab:dapp_boundaries}, the builder-based identification method can ensure that contracts are not misclassified while ensuring high accuracy (i.e., 89.72\%).

\para{Effectiveness of transaction-based replay}
To test the effectiveness of transaction-based replay, we compare the original execution results of all transactions, currently a total of 678,380, with the results obtained through replay for contracts in $D_{unknown}$. Finally, 659,196 transactions, around 97\%, have consistent results. After analysis, we identify the main causes of inconsistency as follows: (1) When multiple transactions in a block invoke the same contract, replaying one of these transactions may lead to inconsistency. For example, if a user deposits tokens and then performs a transfer in two external transactions, directly replaying the transfer transaction may result in a revert, as the account's balance might be insufficient. (2) To test more transactions, \system{} does not consider the gas costs. Thus, \system{} may successfully execute transactions that revert due to insufficient gas.

\para{Effectiveness of cross-DApp analysis}
To validate the effectiveness of cross-DApp analysis, we remove the cross-DApp analysis module (without CDA). More specifically, we replace all cross-DApp analysis in \system{} with traditional cross-contract analysis and \system{} will consider all the contracts encountered in the execution as unsafe. The experimental result is shown in Fig.~\ref{fig:compare_result}. Although without CDA successfully reports the same amounts of true positives as \system{} does, the failure to recognize the boundaries between contracts results in reporting an excessive number of false positives. Besides, without CDA needs to test all contracts blindly, leading to more time costs (Section~\ref{subsec:efficiency}).

\para{Effectiveness of multi-function fuzzing method}
To validate the effectiveness of the multi-function fuzzing method, we replace it with random fuzzing (without MFF). In other words, for each potential entry function, \system{} attempts to generate input based on the original transaction and the function's ABI to simulate a user interacting with two DApps simultaneously. The result is shown in Fig.~\ref{fig:compare_result}. Due to most users not having states in both DApps, such as holding assets in both DApps, most transactions revert as they cannot pass the internal check in both functions or cannot find the critical path of ROR, resulting in a significant amount of false negatives.

\begin{tcolorbox}[colback=SeaGreen!10!CornflowerBlue!10,colframe=RoyalPurple!55!Aquamarine!100!, title = {Answer to RQ 2:}]
% \small
Each module of \system{} indeed helps reduce false positives and false negatives, and improves detection effectiveness.
\end{tcolorbox}

\subsection{Comparison with other Tools}
\label{subsec:with_other_tool}
To answer RQ3, we use $D_{labeled}$ to test with the two most advanced tools, ityFuzz~\cite{ityFuzz} and Sailfish~\cite{sailfish}, as they are currently the most effective dynamic analysis tool and static analysis tool for detecting reentrancy vulnerabilities. We obtain the publicly released artifacts of these two tools from their respective publications. Note that, we do not compare \system{} with other popular tools such as icyChecker~\cite{icychecker}, which can only detect reentrancy within DApps and is not aligned with our scope. Besides, Mythril~\cite{mythril}, Oyente~\cite{oyente}, and Vandal~\cite{Vandal} have been proven less effective than Sailfish in detecting reentrancy~\cite{sailfish}. We conduct the tests using the default configurations of these tools with a time limit of five minutes and one hour for Sailfish and ityFuzz respectively and manually identify the results as \system{} does. After manual inspection, the results are presented in Fig.~\ref{fig:compare_result}. We can observe that ityFuzz and Sailfish are unable to detect any RORs.

Our further investigation finds that although ityFuzz has made optimizations in the fuzzing process, such as using on-chain states to emulate the real environment and using waypoints to improve input quality, it is not effective when it comes to ROR, as ROR requires generating multiple function inputs simultaneously. Besides, Sailfish is a static analysis tool that lacks information about the execution context, such as the specific address called by opcode CALL. Consequently, its call chain is incomplete and cannot detect RORs. Additionally, both tools randomly select functions for testing, which results in lower efficiency compared to \system{}.

\begin{tcolorbox}[colback=SeaGreen!10!CornflowerBlue!10,colframe=RoyalPurple!55!Aquamarine!100!, title = {Answer to RQ 3:}]
% \small
Compared to advanced tools, \system{} detects ROR with higher precision and recall.
\end{tcolorbox}

\begin{table}[htbp]
\centering
\caption{Overall effectiveness of \system{} on $D_{unknown}$.}
\label{tab:resultOfTest}
\small
% \resizebox{0.48\textwidth}{7mm}{
% \begin{tabular}{ccccc}
% \hline
% \multirow{2}{*}{Total Smart Contracts} & \multirow{2}{*}{Reported RORs} & \multicolumn{3}{c}{Detection results} \\ \cline{3-5} 
%                        &                           & TP      & FP      & Precision     \\ \hline
% 2,676                      & 55                         & 50         & 5         & 90.91\%      \\ \hline      
% \end{tabular}
% }
\scalebox{0.95}{
\begin{tabular}{ccccc}
\hline
\multirow{2}{*}{Total Smart Contracts} & \multirow{2}{*}{Reported RORs} & \multicolumn{3}{c}{Detection results} \\ \cline{3-5} 
                       &                           & TP      & FP      & Precision     \\ \hline
2,676                      & 55                         & 50         & 5         & 90.91\%      \\ \hline      
\end{tabular}
}
\end{table}

% To this end, we can conclude that \system{} is indeed more effective than existing tools in detecting ROR.

\subsection{Large-scale Analysis for Finding Unknown RORs}
\label{subsec:large_scale}
To answer RQ4, we evaluate \system{} based on $D_{unknown}$ and Table~\ref{tab:resultOfTest} shows the detailed results. From Table~\ref{tab:resultOfTest} we can find \system{} reports 55 RORs. To validate the results detected by \system{}, our three domain experts independently verified the detection results, and 50 out of the 55 vulnerabilities are confirmed by the experts and the overall precision is 90.91\%. In addition, out of these 50 RORs, we discover that 43 RORs have not been publicly reported before and we find 35 new functions that are suffered with RORs. The total asset affected by these RORs is around 520,000 USD. We will analyze a case in Section~\ref{subsec:case_study} that can only be detected by the \system{} but not by other advanced tools.

\begin{tcolorbox}[colback=SeaGreen!10!CornflowerBlue!10,colframe=RoyalPurple!55!Aquamarine!100!, title = {Answer to RQ 4:}]
% \small

\system{} is effective in RORs under complex DApp interactions in real-world scenarios.

\end{tcolorbox}
% \system{} achieves an accuracy of 90.91\% on a large-scale dataset, successfully detecting 43 unknown RORs and 35 new vulnerable functions, and can protect a substantial amount of funds, further validating the effectiveness of the \system{} on protecting real-world smart contracts.

\begin{table}[htbp]
\centering
    \caption{Performance of \system{} and without CDA on $D_{labeled}$.}
    \label{tab:efficiency}
    \small
    % \resizebox{0.49\textwidth}{9.7mm}{
    % \begin{tabular}{c|cc|c}
    % \hline
    % \multirow{2}{*}{} & \multicolumn{2}{c|}{Avg. Time (seconds)}                     & \multirow{2}{*}{Execution Count} \\ \cline{2-3}
    %                   & \multicolumn{1}{c|}{Per Execution} & Finding ROR &                                  \\ \hline
    % SmartReco         & \multicolumn{1}{c|}{4.12}               & 265.84             & 150421                           \\ \hline
    % without CDA       & \multicolumn{1}{c|}{5.32}               & 485.26             & 558438                           \\ \hline
    % \end{tabular}
    % }
    \scalebox{0.9}{
    \begin{tabular}{c|cc|c}
    \hline
    \multirow{2}{*}{} & \multicolumn{2}{c|}{Avg. Time (seconds)}                     & \multirow{2}{*}{Execution Count} \\ \cline{2-3}
                      & \multicolumn{1}{c|}{Per Execution} & Finding ROR &                                  \\ \hline
    SmartReco         & \multicolumn{1}{c|}{4.12}               & 265.84             & 150421                           \\ \hline
    without CDA       & \multicolumn{1}{c|}{5.32}               & 485.26             & 558438                           \\ \hline
    \end{tabular}
    }

\end{table}

\subsection{Efficiency of \system{}}
\label{subsec:efficiency}
To answer RQ5, we use $D_{labeled}$ to compare the efficiency of \system{} and without CDA in terms of average time per execution, the average time to detect ROR, and number of executions. The results are shown in Table~\ref{tab:efficiency}. Note that we do not compare with without MFF, ityFuzz, and Sailfish, because they do not detect any ROR. Due to the caching mechanism, although \system{} fetches on-chain data, there is no need to repeatedly retrieve after the initial acquisition. Therefore, the average execution time of \system{} is not high. Although the average time per execution of without CDA and \system{} are similar, without CDA has a larger search space and needs to execute more times, resulting in lower efficiency in detecting ROR compared to \system{}. To this end, the performance of \system{} will not be affected by the size of the dataset. Specifically, since the cross-DApp analysis can effectively filter the search space, a larger dataset only leads to at most a linear growth in the number of functions and transactions that need to be analyzed.

\begin{tcolorbox}[colback=SeaGreen!10!CornflowerBlue!10,colframe=RoyalPurple!55!Aquamarine!100!, title = {Answer to RQ 5:}]
% \small
\system{} is rather efficient in detecting ROR with large-scale analysis.
%\system{} can effectively detect ROR. Besides, the cross-DApp analysis used by \system{} is very effective in improving detection efficiency.
\end{tcolorbox}

\subsection{Case Study}
\label{subsec:case_study}
Fig.~\ref{casestudy} is a real case detected by \system{}. In detail, \textit{joinPool} is the entry point for becoming a member of entry DApp. Normally, users need to synchronously transfer a certain amount of funds, like ETH, as collateral. Besides, \textit{joinPool} provides a protection mechanism, line 5 in Fig.~\ref{casestudy}(\subref{casestudy2}), to ensure that users do not transfer funds than expected. To exploit ROR in Fig.~\ref{casestudy}, the first step is to trigger line 5 and invoke line 7 in Fig.~\ref{casestudy}(\subref{casestudy2}). After obtaining control of the transaction, attackers can obtain more funds through \textit{removeLiquity} in Fig.~\ref{casestudy}(\subref{casestudy1}). However, normal transactions are unlikely to trigger this path, as this will incur higher costs. We analyze all the transactions of \textit{joinPool} in contract \textit{Vault} and find that none of them trigger this mechanism. Therefore, solely relying on transactions for testing can result in potential false negatives. \system{} can discover that \textit{joinPool} is payable and will try to mutate the funds carried by \textit{joinPool}, ultimately triggering the protection mechanism and finding the ROR.

\begin{figure}
  \begin{subfigure}[t]{0.47\textwidth}
    \begin{lstlisting}[language=Solidity]
contract Periphery{
  function removeLiquity() public nonReentrant{
    ...
    // get outdated balance
    uint balance = Vault.getBalance();
    IAsset asset = convertERC20ToAssets(balance);
    withdrawTokens(assets);
  }
}
    \end{lstlisting}
    \caption{The victim DApp contract.}
    \label{casestudy1}
  \end{subfigure}
  \begin{subfigure}[b]{0.47\textwidth}
    \begin{lstlisting}[language=Solidity]
contract Vault{
  function joinPool(address pool) public payable nonReentrant {
    ...
    _processJoinPoolTransfers();
    if (balanceTotal - balanceUsed > 0) {
        // root entry point of ROR
        _handleRemainingBalance();
    }
    updatePoolBalance();
    ...
  }
  // The manipulable function 
  function getBalance() view{
    return balance;
  }
}
    \end{lstlisting}
    \caption{The entry DApp contract.}
    \label{casestudy2}
  \end{subfigure}
  \caption{Simplified real-world smart contracts with ROR.}
  \label{casestudy}
\end{figure}

% \system{} mutates transactions of \textit{joinPool} and combines them with original transactions of \textit{removeLiquidity} to detect ROR. Existing tools like ityFuzz struggle to pass the internal check in both functions due to their random input generation. As a result, ityFuzz can not find feasible paths of ROR. Sailfish, on the other hand, is unable to obtain contextual information during execution, such as the specific calling address \textit{adapter}, which prevents it from finding contract \textit{Vault}. An interesting aspect of this example is that the vulnerability can only be triggered by mutating balances of \textit{joinPool}. After analyzing, we discover that normal transactions are unlikely to intentionally follow uncommon paths, such as line 8 in Fig.~\ref{casestudy}(\subref{casestudy2}), as this will incur higher costs. Therefore, solely relying on combining two transactions for testing can result in potential false negatives.

\section{Discussion}
\label{sec:discussion}
\subsection{Detection on deployed contracts}
% \para{Detection on deployed contracts}
\system{} primarily targets the deployed, on-chain contracts, as simulating real DApp interactions, such as obtaining a user's balance or a token price, in an off-chain environment is difficult. In the meantime, we would like to note that for DApps and their smart contracts before deployment, \system{} can utilize existing united test suites or fuzzers to generate transactions and perform detection, similar to existing research~\cite {liu2022finding}. Additionally, detecting deployed contracts is with high-value, as it can help the DApp owners find risks in advance and take measures to reduce losses. For example, deploying a new contract and transferring the funds from the old contract to the new one.

% ROR can only occur during interactions between DApps. Besides, simulating real DApp interactions, such as obtaining a user's balance or a token price, in an off-chain environment is difficult. Therefore, \system{} primarily targets the detection of deployed contracts, but if a complete local environment is provided, \system{} can also perform detection. Additionally, detecting deployed contracts is equally valuable, as it can help deployed contracts find risks in advance and take measures to reduce losses, such as redeploying a new contract and transferring the funds from the old contract to the new one.

% \para{Mitigation against ROR}
% Due to the high complexity of ROR, verifying ROR through existing tools or manual auditing is difficult. To avoid ROR, developers should adhere to the Checks-Effects-Interactions pattern~\cite{wohrer2018smart}, which involves checking the contract to interact, updating the state, and then performing the interaction. 
% Another approach is to add reentrancy checks, such as \textit{nonReentrant}, to all publicly accessible functions within the contract. However, this method incurs more costs. Additionally, DApps should carefully check other involving DApps to avoid potential cross-DApp attacks. 

\subsection{Threats to Validity}
\label{subsec:threats}
\para{Internal threats}
The internal threats of \system{} mainly come from its dependence on clear DApp boundaries. The effectiveness of \system{} could be affected if it fails to accurately recognize the boundary of a given DApp. However, for contracts with unclear boundaries, \system{} will perform analysis on all unknown contracts.  As the results are shown in Section~\ref{subsec:effeciveness_of_each_module}, although \system{} may miss some contract DApp information, there will be no misclassification, ensuring the effectiveness. 
Another major internal threat comes from the evaluation over a limited number of DApps (i.e., 123 popular DApps). Since \system{} needs to test multiple functions simultaneously, the time overhead is larger than traditional single-contract fuzzing tools, which has limited our ability to conduct larger-scale evaluations. However, compared to randomly selected DApps, the DApps in our dataset are representative as they are top-popular DApps. 
%which strengthens the credibility of our experiments.

\para{External threats}
The external threats of \system{} mainly come from the inability to analyze non-Solidity languages, such as Vyper, and contracts without source code. This is because \system{} relies on Slither, which is based on Solidity source code for analysis (although the latest version of Slither claims to cover Vyper, it currently only supports version 0.3.7~\cite{slither_vyper}). We consider this to have a relatively small impact on the effectiveness of \system{}. On the one hand, Solidity is currently the most mainstream smart contract development language~\cite{solidity_is_better}. On the other hand, most DApps tend to open-source code~\cite{zhang2024nyx}, as users are more willing to invest in open-source DApps. To this end, \system{} only misses a small portion of contracts.

% , and targeting Solidity can effectively protect the development of the DApp ecosystem

% Admittedly, \system{} has the following limitations which can be further improved. \jw{\system{} needs to improve the inference ability of the DApp boundaries identification method. Specifically, for an unknown builder, \system{} can infer which DApp it belongs to based on its historical transactions. For example, if the builder interacts frequently with a DApp's contracts, it is likely to belong to that DApp. To reduce potential false negatives, \system{} needs to provide support for other smart contract languages, such as Vyper. For example, parsing the AST generated by Vyper compilation to perform analysis. To reduce false positives, \system{} should increase support for intra-function checks to lower the potential for misidentifying potential entry functions.}

\section{Related Work}
\label{chap:related_work}
\para{Vulnerability detection in smart contract}
At present, most research uses static and dynamic analysis to detect vulnerabilities in smart contracts. Static analysis detects vulnerabilities by analyzing the source code\cite{slither,zeus,smartcheck} or bytecode~\cite{ethainter,defiTainter,achecker} of contracts. For example, DeFiTainter uses decompiled bytecode to detect price manipulation. Slither translates source code into an intermediate representation, called SlitherIR, and performs analysis on it. Due to their inability to obtain execution context information, they miss critical information and cannot fully recover the call graph when detecting ROR. On the other hand, dynamic analysis technologies try to generate inputs that satisfy specific requirements, such as covering more branches. Then by executing the inputs, they can monitor the runtime status of the program and detect vulnerabilities~\cite{contractFuzzer,confuzzius,su2022effectively,reguard}. However, the lack of effective input generation methods in existing tools leads to a significant number of false negatives in ROR, as it is difficult for them to bypass the internal check in functions of several DApps.

\para{Reentrancy in smart contract}
Reentrancy is one of the most common vulnerabilities in smart contracts, and in recent years, many studies have focused on reentrancy~\cite{zhang2022reentrancy,oyente,smartian,sailfish,ityFuzz}. For example, ReVulDL uses deep learning methods to detect reentrancy. ityFuzz employs online fuzzing methods to provide realistic testing. Clairvoyance~\cite{Clairvoyance} uses path-protective techniques to identify potential paths for detecting reentrancy. icyChecker~\cite{icychecker} uses replay to find out potentially vulnerable functions. However, existing works on reentrancy appear to have very limited capability~\cite{reentrancySurvey}, as they overlook the potential correlations between contracts, such as using DApp information to detect vulnerabilities. Unlike existing approaches, \system{} incorporates cross-DApp analysis to obtain more fine-grained information and uses a multi-function fuzzing method to effectively detect ROR, a novel type of reentrancy vulnerability.

\section{Conlucsion}
\label{chap:conclusion}
In this paper, we focus on the detection of ROR based on fine-grained cross-DApp analysis. We introduce \system{}, a novel detection framework to detect RORs. Specifically, we propose a new DApp identification method to identify DApp boundaries from smart contracts. Additionally, \system{} collects DApp-based contextual data based on replay, and uses this information to find potential entry functions. Then, \system{} employs a multi-function fuzzing method to detect RORs. Experimental results show that \system{} achieves good effectiveness, outperforming existing advanced tools. Besides, \system{} successfully detects 43 RORs that have never been reported in real-world smart contracts. 
%The total amount of funds affected by these RORs is approximately 520,000 USD.
% The experimental results over the ground-truth dataset and the large-scale dataset indicate that \system{} outperforms the existing tools in detecting ROR. In total, \system{} successfully finds 5 unreported RORs and 14 contracts have more than 2000 transaction amounts
% By replaying transactions and leveraging DApp information, we obtain crucial execution data such as the DApp information and state modification. Through both intra-DApp and inter-DApp analysis, we determine potential entry functions. Finally, we employ a multi-function fuzzing method to detect ROR vulnerabilities.

\section*{Acknowledgment}

This research is supported by the National Key Research and Development Program of China (2023YFB2704801), NSFC/RGC Collaborative Research (62461160332, CRS\_HKUST602/24), the Major Key Project of PCL (Grant No. PCL2023A05).

\bibliographystyle{IEEEtran}
\bibliography{reference}
\end{document}